\documentclass[aps, prd, twocolumn, superscriptaddress, nofootinbib, amsmath, amssymb]{revtex4-1}

\usepackage{graphicx}  
\usepackage{mathrsfs} 
\usepackage[normalem]{ulem} 

\begin{document}

\title{$\theta$ dependence in trace deformed $SU(3)$ Yang-Mills theory: a lattice study}

\author{Claudio Bonati}
\email{claudio.bonati@df.unipi.it}
\affiliation{Dipartimento di Fisica dell'Universit\`a di Pisa and INFN
  - Sezione di Pisa, Largo Pontecorvo 3, I-56127 Pisa, Italy.}

\author{Marco Cardinali}
\email{marco.cardinali94@yahoo.it}
\affiliation{Dipartimento di Fisica dell'Universit\`a di Pisa and INFN
  - Sezione di Pisa, Largo Pontecorvo 3, I-56127 Pisa, Italy.}

\author{Massimo D'Elia}
\email{massimo.delia@unipi.it}
\affiliation{Dipartimento di Fisica dell'Universit\`a di Pisa and INFN
  - Sezione di Pisa, Largo Pontecorvo 3, I-56127 Pisa, Italy.}

\begin{abstract}
In this paper we investigate, by means of numerical lattice simulations, the
topological properties of the trace deformed $SU(3)$ Yang-Mills theory defined
on $S_1\times\mathbb{R}^3$. More precisely, we evaluate the topological
susceptibility and the $b_2$ coefficient (related to the fourth cumulant of the
topological charge distribution) of this theory for different values of the
lattice spacing and  of the compactification radius. In all the cases we find
results in good agreement with the corresponding ones of the standard $SU(3)$
Yang-Mills theory on $\mathbb{R}^4$.
\end{abstract}

\maketitle

\section{Introduction} 

The strongly interacting dynamics of nonabelian gauge
theories at low energy eluded so far any first-principle analytical
description, although several nonperturbative approximation schemes have been
developed during the years in order to improve our analytical control over this
problem, like the expansion in the number of colors $N_c$ or in the number of
flavours $N_f/N_c$ \cite{ColemanBook, Lucini:2012gg}, instanton calculus
\cite{ColemanBook, Schafer:1996wv} and holographic approaches
\cite{Aharony:1999ti}, just to name a few.  These approaches gave invaluable
hints and helped in clarifying some aspects of the strongly interacting theory,
however they typically provide only qualitative or semi-quantitative results.
Reliable quantitative estimates can still be obtained only numerically, by
means of lattice simulations, or by using effective theories that encode from
the beginning some nonperturbative features, like chiral perturbation theory.

A complementary strategy that has been proposed consists in deforming the
original theory in such a way as to drive the dynamics towards tractable
regimes. For this strategy to be usable one has to ensure that physical
observables are analytic in the deformation, in order to have the possibility
of going back smoothly to the original non deformed case once results have been
obtained in the deformed theory. 

One of the first possibility that may come to mind is to introduce a new scale
in the theory by changing the topology of the space-time from $\mathbb{R}^4$ to
$S_1\times \mathbb{R}^3$, where $S_1$ stands for a circumference of length $L$.
By varying $L$ we switch between the original theory on $\mathbb{R}^4$ (case
$L\gg \Lambda^{-1}$, with $\Lambda$ a typical energy scale of the theory) and a
regime in which perturbation theory and instanton calculus can be applied (case
$L\ll \Lambda^{-1}$). 

What remains to be shown, in order to advocate the compactification on
$S_1\times\mathbb{R}^3$ as useful in this paradigm, is that physical properties
change smoothly when varying the compactificaton 
{length} $L$. This is however generically not the case: the
compactified theory resembles very much (and for some choice of boundary
conditions it is) finite temperature field theory, and a phase transition is
likely to happen at finite temperature.

From now on we will consider the specific case of $SU(3)$ Yang-Mills theory
compactified on $S_1\times\mathbb{R}^3$ with periodic boundary conditions. In
this setup the {length of the compactified
direction} is nothing but the inverse temperature and it is well known that for
$L\approx 0.7~\mathrm{fm}$ (corresponding to a temperature $T_c \simeq
270~\mathrm{MeV}$) a first order phase transition is present
\cite{Boyd:1996bx}, separating the low temperature confined phase from the high
temperature deconfined one.  It is clear that in such a situation it is
hopeless to obtain reliable results for the large $L$ case by studying the
small $L$ case. To proceed further with this approach we have to smoothen or
remove the phase transition and here the trace deformation of the action
enters.

Let us remind the reader that the deconfinement phase transition at finite
temperature is associated with the spontaneous symmetry breaking (SSB) of the
$Z_3$ center symmetry, whose order parameter is the mean value of the trace of
the Polyakov loop $P(\vec{x}) = \mathcal{P}\exp\left(i\int_0^{L}A_0(\vec{x}, t)
\mathrm{d}t\right)$, which vanishes in the confined phase ($\langle
\mathrm{Tr}P\rangle =0$) while it is different from zero 
for $T > T_c$
($\langle \mathrm{Tr}P\rangle=\alpha e^{i 2\pi n/3}$, with $n\in\{0, 1,
2\}$ and $\alpha>0$).  

In order to remove the $Z_3$ SSB that prevents a smooth connection between
large and small $L$ regimes, it was suggested in \cite{Unsal:2008ch} to add to
the $SU(3)$ Yang-Mills action a new term, explicitly dependent on the Polyakov
loop and disfavouring configurations with $\mathrm{Tr}P\neq 0$ in the
path-integral.  Inspired by the perturbative form of the Polyakov loop
effective action at high temperature \cite{Gross:1980br}, the authors of
\cite{Unsal:2008ch} suggested the following form for the new term:
\begin{equation}\label{eq:deform_cont}
S_{\mathrm{td}}=h\int |\mathrm{Tr}P(\vec{x})|^2\mathrm{d}^3 x\ ,
\end{equation}
where $h$ is a new parameter and the subscript ``td'' stands for ``trace
deformation'' (higher powers of $P(\vec{x})$ have also to be added in $SU(N_c)$
theories with $N_c>3$, see \cite{Unsal:2008ch}). Several works followed this
approach, but possible alternative, like the introduction of adjoint fermions
or the use of non-thermal boundary conditions, have also been proposed
\cite{Kovtun:2007py, Unsal:2007vu, Unsal:2007jx, Shifman:2008ja, Myers:2009df,
Cossu:2009sq, Meisinger:2009ne, Unsal:2010qh, Thomas:2012ib, Poppitz:2012sw,
Thomas:2012tu, Poppitz:2012nz, Misumi:2014raa, Anber:2014lba, Bhoonah:2014gpa,
Cherman:2016vpt, Sulejmanpasic:2016llc, Anber:2017rch}. In the present work we
will restrict ourselves to the case of the deformation in
Eq.~\eqref{eq:deform_cont}.

It has been shown in \cite{Myers:2007vc}, using numerical lattice simulations,
that the new term $S_{\mathrm{td}}$ indeed moves to smaller values the critical
compactification {length} at which deconfinement happens,
but it also introduces a new phase (called ``skewed'') that has no equivalent
in the non deformed theory. A systematic study of the changes induced by
$S_{\mathrm{td}}$ on observables different from $\langle \mathrm{Tr}P\rangle$
has however never been undertaken so far and the present work is a first step
in this direction.

The reason for performing such a study is that there is no way of excluding
\emph{a priori} the possibility that the deformation term $S_{\mathrm{td}}$
generates some spurious phase transition in observables uncorrelated with
center symmetry. From a more general perspective we can ask: are we sure that
what really matters in the low energy dynamics of $SU(3)$ Yang-Mills is
\emph{just} the fact that center symmetry in not spontaneously broken? Since we
have no definite answer to this fundamental question, the best thing we can do
is to study the trace deformed theory by means of lattice simulations and
investigate the behavior of not-center-related physical observables as
functions of $h$.

In the present work we concentrate on two observables related to $\theta$
dependence: the topological susceptibility $\chi$ and the coefficient $b_2$,
related to the fourth order cumulant 
of the topological charge distribution
(see, e.g., \cite{Vicari:2008jw}). These observables appear to be perfectly suited to
our purposes, since their value is fixed only by non-perturbative physics, they
are very sensitive to the deconfinement transition \cite{Alles:1997qe,
Gattringer:2002mr, Lucini:2004yh, DelDebbio:2004vxo, Bonati:2013tt} and they do
not appear to be tightly related to center symmetry \cite{Ilgenfritz:2012wg,
Bonati:2015uga}.

\section{Numerical setup} 

The standard Wilson action \cite{Wilson:1974sk} with
bare coupling $\beta=6/g^2$ was used to discretize the theory and the addition
of the term $S_\mathrm{td}$ presents no difficulties, but for the
fact that now the action is nonlinear in the temporal links. For this reason a
simple Metropolis scheme \cite{Metropolis:1953am} had to be used to update
temporal links, while links directed along other directions could be
updated by heatbath and overrelaxation algorithms \cite{Creutz:1980zw,
Kennedy:1985nu, Creutz:1987xi} implemented \emph{\`a la} Cabibbo-Marinari
\cite{Cabibbo:1982zn}. 

To measure the topological content of the gauge configurations we used
cooling~\cite{Berg:1981nw, Iwasaki:1983bv, Itoh:1984pr, Teper:1985rb,
Ilgenfritz:1985dz} to remove fluctuations at the scale of the lattice spacing
(see \cite{Bonati:2014tqa, Cichy:2014qta, Namekawa:2015wua, Alexandrou:2015yba,
Berg:2016wfw, Alexandrou:2017hqw} for discussions on the practical equivalence
of different smoothing algorithms) and we measured the topological charge
$Q=\int q(x)\mathrm{d}^4x$ on the smoothed configurations using the
discretization of $q(x)$ introduced in \cite{DiVecchia:1981qi, DiVecchia:1981hh}
\begin{equation}\label{eq:qlattice}
q_L(x) = -\frac{1}{2^9 \pi^2} 
\sum_{\mu\nu\rho\sigma = \pm 1}^{\pm 4} 
{\tilde{\epsilon}}_{\mu\nu\rho\sigma} \hbox{Tr} \left( 
\Pi_{\mu\nu}(x) \Pi_{\rho\sigma}(x) \right) \; ,
\end{equation}
where $\Pi_{\mu\nu}$ denotes the plaquette operator,
$\tilde{\epsilon}_{\mu\nu\rho\sigma}$ coincides with the Levi-Civita tensor for
positive entries and  
is fixed by complete
antisymmetry and ${\tilde{\epsilon}}_{\mu\nu\rho\sigma} =
-{\tilde{\epsilon}}_{(-\mu)\nu\rho\sigma}$ otherwise.

The topological susceptibility $\chi$ and the $b_2$ coefficient parameterize 
up to $O(\theta^4)$ the $\theta$ dependence of the vacuum energy density 
\cite{Vicari:2008jw} 
\begin{equation}\label{eq:theta_dep}
\Delta E(\theta) \equiv E(\theta)-E(0)=\frac{1}{2}\chi\theta^2(1+b_2\theta^2+b_4\theta^4+\cdots)
\end{equation}
and they can be related to the cumulants of the topological charge distribution
at $\theta=0$ by the relations \cite{Vicari:2008jw} 
\begin{equation}\label{eq:chi_b2}
\chi = \frac{\langle Q^2 \rangle_{\theta=0}}{\mathcal{V}}\ ,\quad
b_2=-\frac{\langle Q^4\rangle_{\theta=0}-3\langle Q^2\rangle^2_{\theta=0}}{12\langle Q^2\rangle_{\theta=0}}\ ,
\end{equation}
where $\mathcal{V}$ is the four-dimensional volume. These expressions can be
used to compute $\chi$ and $b_2$ using simulations performed at $\theta=0$.

\begin{figure}[t]
\includegraphics[width=0.9\columnwidth, clip]{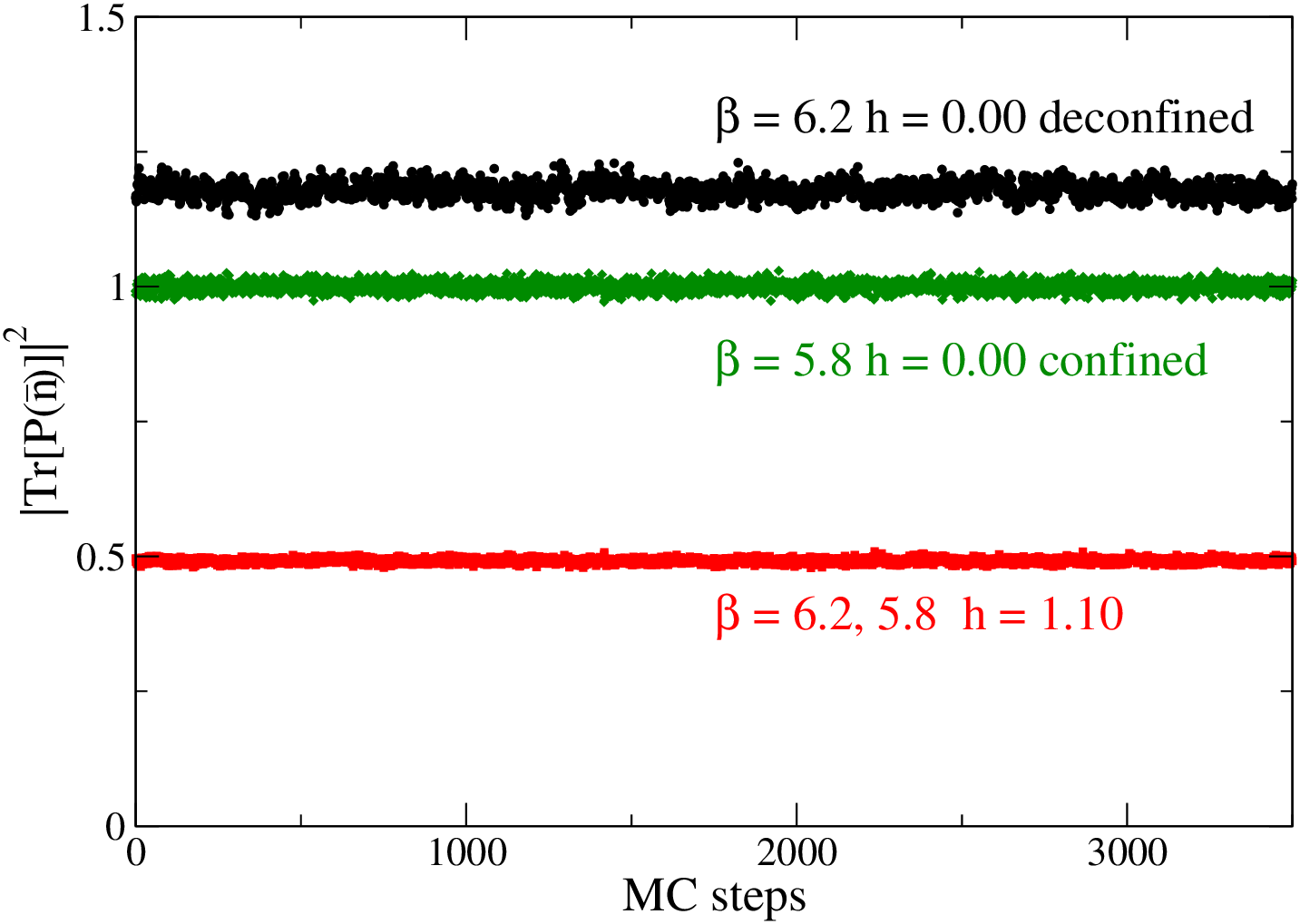}
\caption{Time histories of $|\mathrm{Tr}P(\vec n)|^2$ 
for two values of the bare
coupling ($\beta=5.8$ and $6.2$) and two values of the deformation parameter
($h=0$ and $1.1$), measured on a $8\times 32^3$ lattice.  The two sets of data
corresponding to $\beta=5.8, 6.2$ at $h=1.1$ are graphically
indistinguishable.}
\label{fig:adj_poly}
\end{figure}

While $\theta=0$ simulations represent the optimal strategy if one is
interested in $\chi$, to determine $b_2$ there is a better possibility:
simulations performed at imaginary values (to avoid the sign problem) of
$\theta$ can be used to obtain a better estimator, with improved
signal-to-noise ratio on large volumes \cite{Panagopoulos:2011rb,
Bonati:2015sqt, Bonati:2016tvi}. In this approach one adds to the discretized
Lagrangian density a term $\mathcal{L}_\theta=-\theta_Lq_L(x)$, where
$\theta_L$ is the lattice $\theta$ parameter (related to its
continuum counterpart by a finite renormalization, $\theta=Z\theta_L$
\cite{Campostrini:1988cy}) and $q_L(x)$ is defined in Eq.~\eqref{eq:qlattice}. The
values of $Z$, $\chi$ and $b_2$ can then be obtained 
by fitting the cumulants
of the distribution of the topological charge extracted from simulations
performed at $\theta_L\neq 0$, i.e.
\begin{equation}\label{eq:cum_theta_L}
\begin{aligned}
& \langle Q \rangle_{\theta_L} =
\mathcal{V}\chi Z \theta_L (1 - 2 b_2 Z^2 \theta_L^2 + \dots)\, , \\
& \langle Q^2 \rangle_{\theta_L}-\langle Q\rangle^2_{\theta_L} =
\mathcal{V}\chi (1 - 6 b_2 Z^2 \theta_L^2 + \dots)\, ,
\end{aligned}
\end{equation} 
see \cite{Bonati:2015sqt} for more details. The first four cumulants of the 
topological charge measured at $\theta_L\neq 0$ were used in this work to provide 
precise estimates of $b_2$.

\section{Results} 

Before presenting our results for $\chi$ and $b_2$ in the
deformed theory, let us make a few comments on the way in which center symmetry
can be realized in Yang-Mills theory and in its deformed counterpart. In
ordinary Yang-Mills theory the fact that $\langle \mathrm{Tr}P\rangle=0$ does
not imply that $\langle |\mathrm{Tr}P (\vec n)|^2\rangle$ has to be ``small'', i.e.
fluctuations of the Polyakov loop are not severely constrained in the confined
region. In the confined phase of the deformed theory at small $L$, where
$\langle \mathrm{Tr}P\rangle=0$ is enforced by the new term in 
Eq.~\eqref{eq:deform_cont}, fluctuations of $\mathrm{Tr}P$ are instead
strongly suppressed.

In Fig.~\ref{fig:adj_poly} we report data for $\langle
|\mathrm{Tr}P (\vec n)|^2\rangle$ (related to the trace of $P$ in the adjoint
representation) measured on a $8\times 32^3$ lattice for two values of the bare
coupling $\beta$ and of the parameter $h$ controlling the deformation. Without
deformation ($h=0$ case) the system is in the confined phase at $\beta=5.8$ but
not at $\beta=6.2$; for $h=1.10$ center symmetry is restored also at
$\beta=6.2$. We see that $\langle |\mathrm{Tr}P (\vec n)|^2\rangle\simeq 1$ in the
standard confined phase ($h=0$) while it gets significantly smaller,
$\langle|\mathrm{Tr}P (\vec n)|^2\rangle \simeq 0.5$, when the deformation is switched on
($h=1.1$). This is a possible indication that the confined phase of the
original and of the deformed theory are different from the dynamical
point of view. Will this difference persist in observables of more direct
physical relevance? To elucidate this point we now describe the results
obtained for the $\theta$ dependence in the two cases. 

\begin{figure}[t]
\includegraphics[width=0.9\columnwidth, clip]{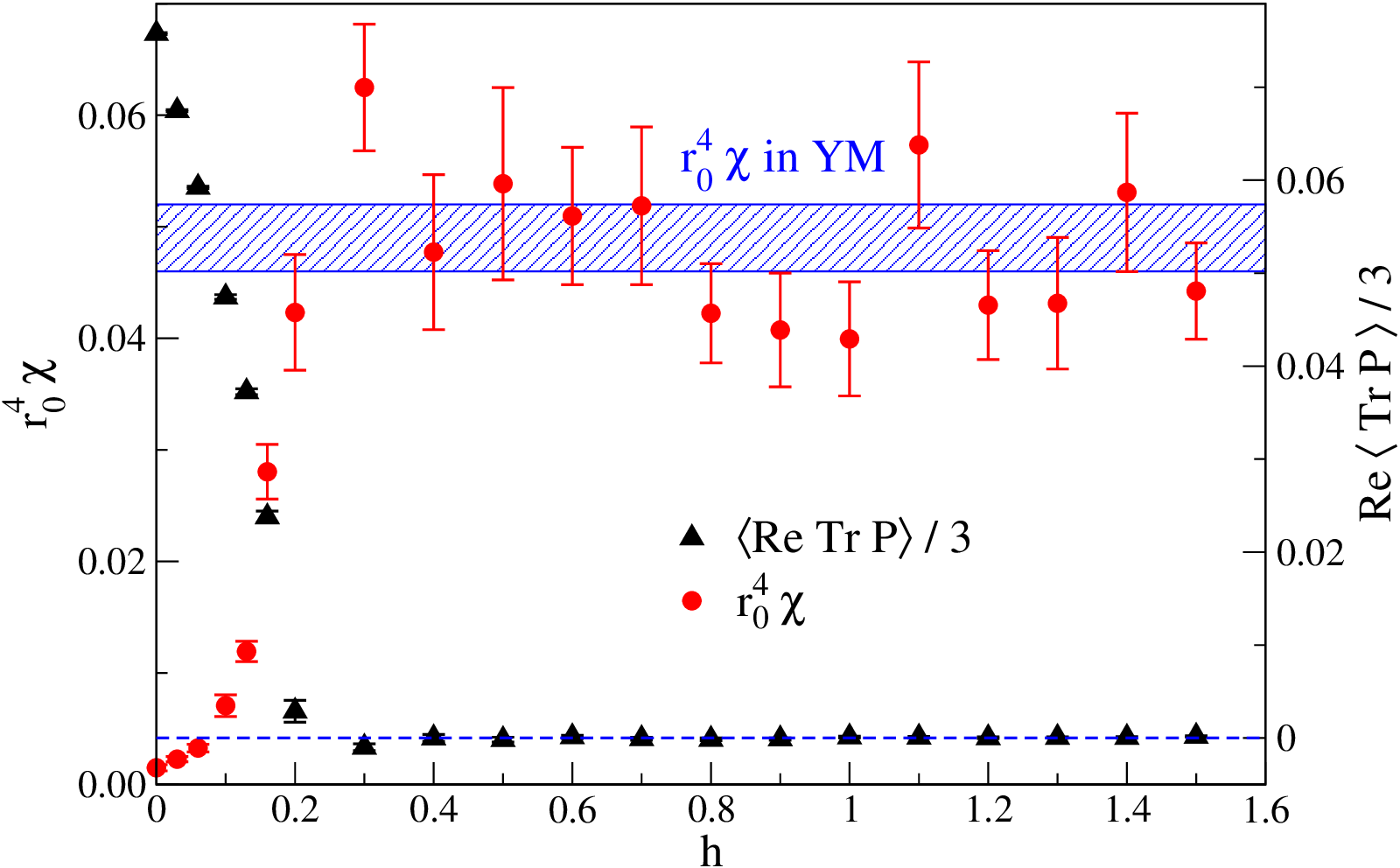}
\caption{Topological susceptibility $\chi$ and $\mathrm{Re}\langle \mathrm{Tr}
P\rangle/3$ measured on a $8\times 32^3$ lattice at bare coupling $\beta=6.4$ as
a function of $h$.  The value obtained in standard $SU(3)$ Yang-Mills theory
\cite{Vicari:2008jw} is also shown for reference (horizontal band).}
\label{fig:r04chi}
\end{figure}

In Fig.~\ref{fig:r04chi} we show the behavior of the topological
susceptibility as a function of $h$, obtained from simulations performed at
$\theta=0$ on an $8\times 32^3$ lattice at coupling $\beta=6.4$. The lattice
spacing is fixed by the value of the Sommer parameter $r_0$
\cite{Sommer:1993ce}, determined in \cite{Guagnelli:1998ud}, whose value in
physical units is $r_0\simeq 0.5~\mathrm{fm}$. For $\beta=6.4$ and temporal
extent $N_t=8$ the system is deconfined at $h=0$ and $\chi$ is very small
\cite{Alles:1997qe, Gattringer:2002mr, Lucini:2004yh, DelDebbio:2004vxo}. By
increasing the value of $h$ the topological susceptibility quickly gets larger,
until it reaches a plateau starting around $h\approx 0.3$, which is
approximately the value at which center symmetry starts to be restored (see the 
behavior of $\mathrm{Re}\langle\mathrm{Tr}P\rangle$ in Fig.~\ref{fig:r04chi}).

\begin{figure}[t]
\includegraphics[width=0.9\columnwidth, clip]{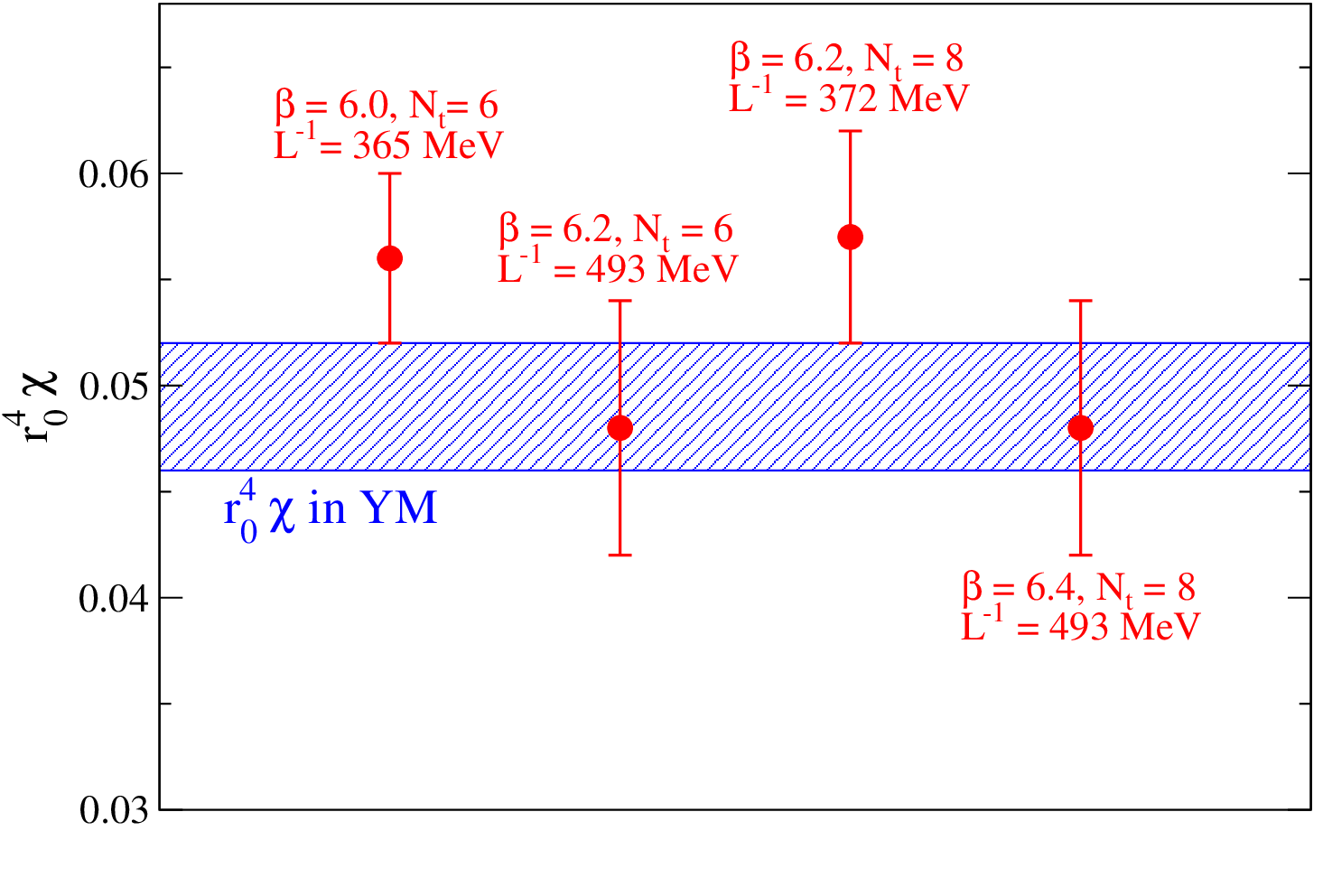}
\caption{Plateau values of $\chi$ extracted from simulations performed on
lattices of different temporal extent ($N_t=6,8$, with $N_s=32$) and using
different couplings ($\beta=6.0, 6.2, 6.4$). 
We also report the inverse compactification size in physical units.}
\label{fig:r04chi_riass}
\end{figure}

In Fig.~\ref{fig:r04chi} the value of $\chi$ obtained in standard $SU(3)$
Yang-Mills theory is also reported for reference and it can be noted that this
value is consistent with that in the plateau region of the deformed theory; the
same happens in all the explored cases. Two different physical values of $L$
have been studied, $L^{-1}\simeq 370\,\mathrm{MeV}$ and $L^{-1}\simeq
495\,\mathrm{MeV}$, and for each of these values two sets of simulations (at
$\theta=0$) have been performed, corresponding to two values of the coupling
$\beta$ and 16 values of $h$ in the range $0\le h \le 1.5$. The qualitative
behavior observed for $\chi$ as a function of $h$ is the same as that shown in
Fig.~\ref{fig:r04chi} and the plateau values are reported in
Fig.~\ref{fig:r04chi_riass}, again together with the standard $SU(3)$ value.
From this figure we can exclude the presence of sizable lattice artefacts in
the $\chi$ plateau values, which are always compatible with the standard
$SU(3)$ value and remarkably insensitive to $L$.

\begin{table}[b]
\begin{tabular}{|l|l|l|}
\hline 
$\beta$ & $h$ & $t_0/a^2$ \\ \hline \hline
5.96    & 0.0   & 2.7854(62)\\ \hline
5.96    & 1.0   & 2.8087(69) \\ \hline
5.96    & 2.0   & 2.8063(74) \\ \hline 
\end{tabular}
\quad 
\begin{tabular}{|l|l|l|}
\hline 
$\beta$ & $h$ & $t_0/a^2$ \\ \hline \hline
6.17    & 0.0   & 5.489(14) \\ \hline
6.17    & 1.0   & 5.530(16) \\ \hline
6.17    & 2.0   & 5.498(16) \\ \hline
\end{tabular}
\caption{Values of $t_0/a^2$ with and without the trace deformation. Values at
$h=0$ have been computed in \cite{Luscher:2010iy}, results at $\beta=5.96$ have
been extracted using $24^4$ lattices, while $32^4$ lattices have been used
at $\beta=6.17$.}
\label{tab:t_h}
\end{table}

Up to now we have tacitly assumed the lattice spacing to be independent of the
deformation parameter $h$. We can improve on this in two different ways: by
explicitly setting the scale at $h\neq 0$ or by looking at dimensionless
observables, whose expectation values are independent of the scale setting.

In order to directly test the independence of the lattice spacing on $h$ we
determined the scale $t_0$, defined by gradient flow and introduced in
\cite{Luscher:2010iy}. While this scale is not associated to the value of a
physical observable of direct experimental relevance (like $r_0$ or the string
tension), it has the advantage of being easily measurable with good accuracy on
the lattice (see e.g. the discussion in \cite{Sommer:2014mea}). To extract the
value of $t_0/a^2$ we integrated the flow equations using the Runge-Kutta
integrator described in App.~C of \cite{Luscher:2010iy} with stepsize
$\epsilon=0.01$, using a statistics of $O(100)$ independent configurations
generated on symmetric lattices. The results obtained are reported in
Tab.~\ref{tab:t_h} and the outcome is that $t_0/a^2$ is indeed practically
independent of $h$ in the expored range: data coincides with those
at $h=0$ up to less than $1\%$, i.e.~well within the statistical
errors on $\chi$.

Finally, let us discuss results for the dimensionless coefficient $b_2$,
defined in Eqs.~\eqref{eq:theta_dep}-\eqref{eq:chi_b2}. As previously
discussed, to obtain precise results for this observable it is convenient to
perform simulations at imaginary values of the $\theta$ parameter, which are
however significantly slower than the $\theta=0$ ones{: a single
simulation is slower than the corresponding one at $\theta=0$ by a factor
$2\div 3$ and several $\theta$ values have to be simulated to extract a
single determination of $b_2$}. For this reason we concentrated on just a
couple of points, well in the plateau region of $\chi$: simulations were
performed for $\beta=6.4$ at two values of the deformation parameter ($h=1.10$
and $1.20$) using $8\times 32^3$ lattices. Seven values of $\theta_L$ (the
lattice imaginary $\theta$ parameter) were investigated, in the range $0\le
\theta\le 16$, and the stability of the results was tested against changes of
the fit range adopted. In all the cases the $O(\theta^6)$ dependence of the
vacuum energy density come out to be negligible (as in ordinary Yang-Mills
\cite{Bonati:2015sqt}) and in the fit we thus used $b_4=0$ (see
Eq.~\eqref{eq:theta_dep}).

\begin{figure}[t]
\includegraphics[width=0.9\columnwidth, clip]{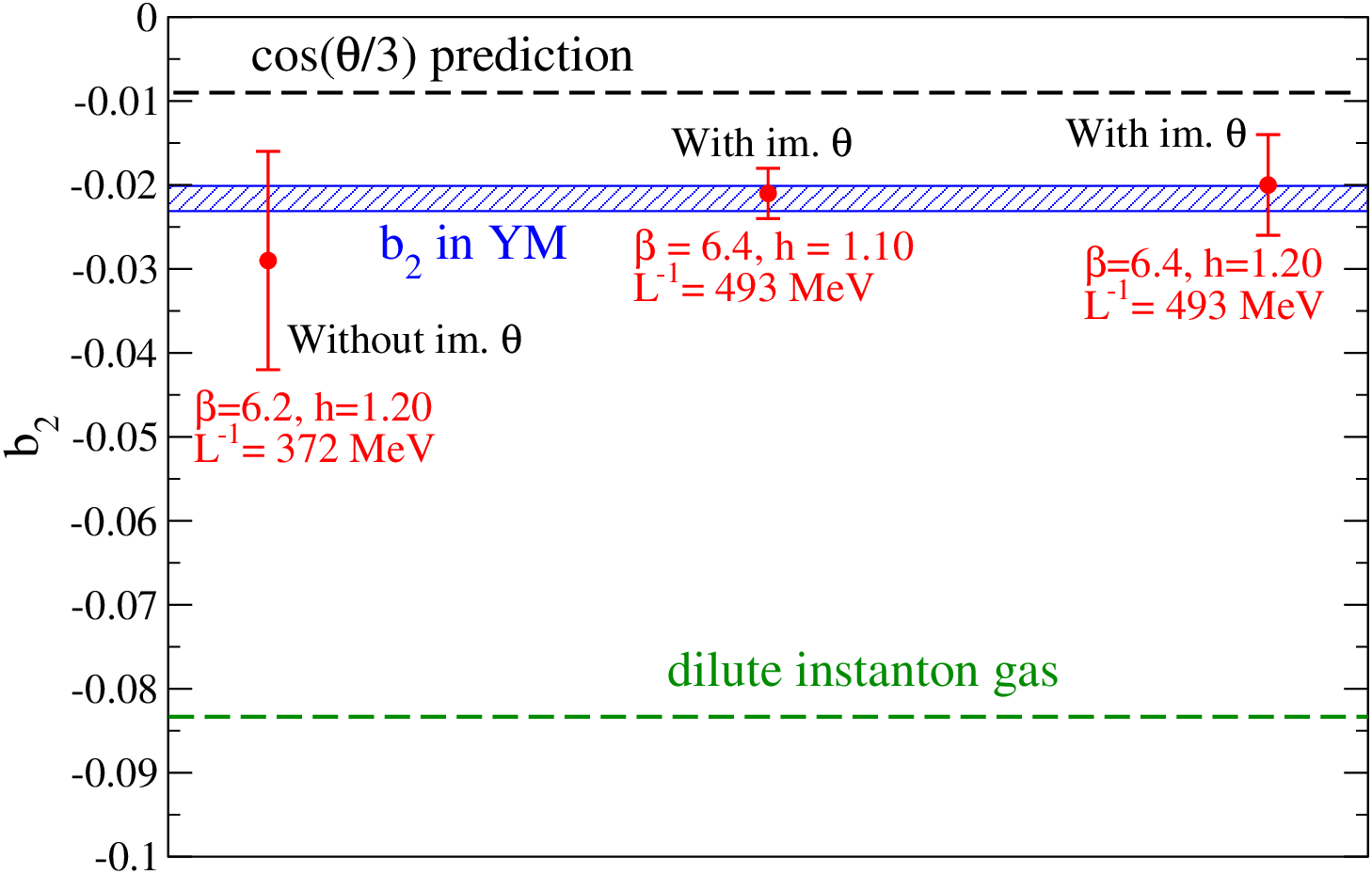}
\caption{Results obtained for $b_2$ in the deformed theory using $8\times 32^3$
lattices. The horizontal band denotes the standard $SU(3)$ value
\cite{Bonati:2015sqt}, dashed lines denote the value $b_2=-1/12$ (DIGA) 
and $b_2= - 1/108$ (Fractional Instanton Gas Approximation).}
\label{fig:b2}
\end{figure}

Results obtained for $b_2$ are shown in Fig.~\ref{fig:b2} together with the
standard $SU(3)$ result of \cite{Bonati:2015sqt}. To appreciate the
effectiveness of the imaginary $\theta$ approach, a point is also shown
obtained by using simulations at $\theta=0$ only, which required about the same
CPU-time as the imaginary $\theta$ ones.  Also for $b_2$ there is very good
agreement between the values at the plateau for the deformed theory and the
values known for the confined Yang-Mills theory~\cite{DelDebbio:2002xa,
DElia:2003zne, Giusti:2007tu, Ce:2015qha, Bonati:2015sqt}, in this case without
any assumption on the lattice spacing, since $b_2$ is dimensionless. 

For comparison, in Fig.~\ref{fig:b2} we also indicate two values of $b_2$
typical of particular regimes. The first is that in which the dominant
topological excitations have integer charges and are weakly interacting. Such a
regime is well described by the dilute instanton gas approximation (DIGA)
\cite{Gross:1980br}, in which $\Delta E(\theta)\propto 1-\cos\theta$ and
$b_2=-1/12$. This value is typical of Yang-Mills theory in the deconfined phase
\cite{Bonati:2013tt} and it is clearly incompatible with the results obtained
in this work. 

Another interesting case is that in which excitations are still
weakly interacting but have fractional topological charges $1/3$ ($1/N_c$ for
$SU(N_c)$).  This regime {corresponds to the functional form $\Delta
E(\theta)\propto1-\cos(\theta/3)$ of the vacuum energy, characterized by
$b_2=-1/108$, and it} is expected to well describe the deformed theory in the
limit of {asymptotically} small $L$ 
values~\cite{Unsal:2008ch, Thomas:2011ee, Aitken:2018mbb}: 
{this sort of fractional instanton gas approximation
is related to the fact that Abelian degrees of freedom are dominant
in the deformed theory in the limit of small $L$}~\cite{Unsal:2008ch, Thomas:2011ee, Aitken:2018mbb}.
 From
Fig.~\ref{fig:b2} we see that our results are inconsistent also with this
value. {A possible interpretation is that, for the explored values
of $L$, the deformed theory resembles the actual Yang-Mills vacuum more closely
than for asymptotically small values of $L$, so that non-Abelian 
degrees of freedom are still relevant, leading to non-trivial interactions
between the fractionally charged objects, hence to a value of $b_2$
which is not equal to the asymptotically predicted one; however, it is also
not far from it, supporting the idea that corrections due to residual interactions might be
small, and maybe analytically computable. Following the same line of reasoning,
based on the virial expansion, 
discussed in Ref.~\cite{Bonati:2013tt}, one might infer that the residual
interactions between the fractionally charged objects are repulsive, because
the deviation of $b_2$ from the asymptotic prediction is negative.}

\section{Conclusions} 

In this paper we investigated, by means of Monte-Carlo
simulations, the non-perturbative dynamics of the trace deformed $SU(3)$ gauge
theory, in which the term in Eq.~\eqref{eq:deform_cont} is added to the action.
Such a deformation term inhibits the spontaneous symmetry breaking of 
center symmetry in the presence of a compactified direction and, in principle,
opens the way to the possibility of investigating the low-energy physics of
Yang-Mills theory using perturbative/semiclassical methods. For such an
ambitious goal to be achievable it is fundamental that physical observables
behave smoothly, as functions of $\beta$ and $h$, up to small values of the
compactification length $L$. In this paper we investigated the behavior of
observables related to the $\theta$-dependence to inquire this point.

Our numerical results for the topological susceptibility and the coefficient
$b_2$, obtained using compactification lengths $L^{-1}\approx
370\,\mathrm{MeV}$ and $L^{-1}\approx 495\,\mathrm{MeV}$, are perfectly
compatible with the known values for the non-deformed $SU(3)$ theory. Given the
completely nonperturbative origin of these quantities, this is a strong
indication that the compactified theory indeed conserves intact a significant
part of the dynamics of the original Yang-Mills theory.

The values obtained for $b_2$ show that, at least for the $L$ values explored,
low-energy physics cannot be described as a gas of weakly interacting objects
of integer or fractional ($1/N_c$) topological charge.  This is again the same
thing that happens in ordinary Yang-Mills, but it is at odds with what is
expected to happen at very small compactification radii in the deformed theory.
{A possible interpretation of this result is the following: the
nonperturbative dynamics of the deformed theory is so similar to that of the
original Yang-Mills one, that analytical computations that go beyond the known
leading order semiclassical approximations are required to quantitatively
describe our numerical data. Indeed the fact that the $L$ dependence is smooth is not enough
to guarantee leading order results to be reliable down to $L\approx 500\,\mathrm{MeV}$.}
This is a point that surely deserves further studies, specifically targeted at
investigating the small $L$ regime {and the way in which the}
large $N_c$ limit {is approached.} {Another interesting
topic that could be relevant to better understand this point is} the nature
of the topological excitations {in the deformed theory}, which have to be
substantially different from that of Yang-Mills theory, {because of the
compactified direction, but} nevertheless with a similar distribution. The
study of other not-$\theta$-related observables  is also something of the
utmost importance to get a complete picture of the physical effects of the
deformation.

\emph{Acknowledgement}
We thank J.~Greensite, M.~Unsal and T.~Sulejmanpasic for useful discussions.
Numerical simulations have been performed at the 
Scientific Computing Center at INFN-PISA 
and on the MARCONI machine at CINECA, based on the 
agreement between INFN and CINECA (under project INF18\_npqcd).


\begin{thebibliography}{95}

\bibitem{ColemanBook} 
  S.~Coleman ``Aspects of symmetry'' Cambridge University Press (1988).
\bibitem{Lucini:2012gg} 
  B.~Lucini and M.~Panero,
  Phys.\ Rept.\  {\bf 526}, 93 (2013)
  [arXiv:1210.4997 [hep-th]].
\bibitem{Schafer:1996wv} 
  T.~Schaefer and E.~V.~Shuryak,
  Rev.\ Mod.\ Phys.\  {\bf 70}, 323 (1998)
  [hep-ph/9610451].
\bibitem{Aharony:1999ti} 
  O.~Aharony, S.~S.~Gubser, J.~M.~Maldacena, H.~Ooguri and Y.~Oz,
  Phys.\ Rept.\  {\bf 323}, 183 (2000)
  [hep-th/9905111].

\bibitem{Boyd:1996bx} 
  G.~Boyd, J.~Engels, F.~Karsch, E.~Laermann, C.~Legeland, M.~Lutgemeier and B.~Petersson,
  Nucl.\ Phys.\ B {\bf 469}, 419 (1996)
  [hep-lat/9602007].

\bibitem{Unsal:2008ch} 
  M.~Unsal and L.~G.~Yaffe,
  Phys.\ Rev.\ D {\bf 78}, 065035 (2008)
  [arXiv:0803.0344 [hep-th]].

\bibitem{Gross:1980br} 
  D.~J.~Gross, R.~D.~Pisarski and L.~G.~Yaffe,
  Rev.\ Mod.\ Phys.\  {\bf 53}, 43 (1981).


\bibitem{Kovtun:2007py} 
  P.~Kovtun, M.~Unsal and L.~G.~Yaffe,
  JHEP {\bf 0706}, 019 (2007)
  [hep-th/0702021 [HEP-TH]].
\bibitem{Unsal:2007vu} 
  M.~Unsal,
  Phys.\ Rev.\ Lett.\  {\bf 100}, 032005 (2008)
  [arXiv:0708.1772 [hep-th]].
\bibitem{Unsal:2007jx} 
  M.~Unsal,
  Phys.\ Rev.\ D {\bf 80}, 065001 (2009)
  [arXiv:0709.3269 [hep-th]].
\bibitem{Shifman:2008ja} 
  M.~Shifman and M.~Unsal,
  Phys.\ Rev.\ D {\bf 78}, 065004 (2008)
  [arXiv:0802.1232 [hep-th]].
\bibitem{Myers:2009df} 
  J.~C.~Myers and M.~C.~Ogilvie,
  JHEP {\bf 0907}, 095 (2009)
  [arXiv:0903.4638 [hep-th]].
\bibitem{Cossu:2009sq} 
  G.~Cossu and M.~D'Elia,
  JHEP {\bf 0907}, 048 (2009)
  [arXiv:0904.1353 [hep-lat]].
\bibitem{Meisinger:2009ne} 
  P.~N.~Meisinger and M.~C.~Ogilvie,
  Phys.\ Rev.\ D {\bf 81}, 025012 (2010)
  [arXiv:0905.3577 [hep-lat]].
\bibitem{Unsal:2010qh} 
  M.~Unsal and L.~G.~Yaffe,
  JHEP {\bf 1008}, 030 (2010)
  [arXiv:1006.2101 [hep-th]].
\bibitem{Thomas:2012ib} 
  E.~Thomas and A.~R.~Zhitnitsky,
  Phys.\ Rev.\ D {\bf 86}, 065029 (2012)
  [arXiv:1203.6073 [hep-ph]].
\bibitem{Poppitz:2012sw} 
  E.~Poppitz, T.~Schaefer and M.~Unsal,
  JHEP {\bf 1210}, 115 (2012)
  [arXiv:1205.0290 [hep-th]].
\bibitem{Thomas:2012tu} 
  E.~Thomas and A.~R.~Zhitnitsky,
  Phys.\ Rev.\ D {\bf 87}, no. 8, 085027 (2013)
  [arXiv:1208.2030 [hep-ph]].
\bibitem{Poppitz:2012nz} 
  E.~Poppitz, T.~Schaefer and M.~Unsal,
  JHEP {\bf 1303}, 087 (2013)
  [arXiv:1212.1238 [hep-th]].
\bibitem{Misumi:2014raa} 
  T.~Misumi and T.~Kanazawa,
  JHEP {\bf 1406}, 181 (2014)
  [arXiv:1405.3113 [hep-ph]].
\bibitem{Anber:2014lba} 
  M.~M.~Anber, E.~Poppitz and B.~Teeple,
  JHEP {\bf 1409}, 040 (2014)
  [arXiv:1406.1199 [hep-th]].
\bibitem{Bhoonah:2014gpa} 
  A.~Bhoonah, E.~Thomas and A.~R.~Zhitnitsky,
  Nucl.\ Phys.\ B {\bf 890}, 30 (2014)
  [arXiv:1407.5121 [hep-ph]].
\bibitem{Cherman:2016vpt} 
  A.~Cherman, S.~Sen, M.~L.~Wagman and L.~G.~Yaffe,
  Phys.\ Rev.\ D {\bf 95}, no. 7, 074512 (2017)
  [arXiv:1612.00403 [hep-lat]].
\bibitem{Sulejmanpasic:2016llc} 
  T.~Sulejmanpasic,
  Phys.\ Rev.\ Lett.\  {\bf 118}, no. 1, 011601 (2017)
  [arXiv:1610.04009 [hep-th]].
\bibitem{Anber:2017rch} 
  M.~M.~Anber and A.~R.~Zhitnitsky,
  Phys.\ Rev.\ D {\bf 96}, no. 7, 074022 (2017)
  [arXiv:1708.07520 [hep-th]].


\bibitem{Myers:2007vc} 
  J.~C.~Myers and M.~C.~Ogilvie,
  Phys.\ Rev.\ D {\bf 77}, 125030 (2008)
  [arXiv:0707.1869 [hep-lat]].

\bibitem{Vicari:2008jw} 
  E.~Vicari and H.~Panagopoulos,
  Phys.\ Rept.\  {\bf 470}, 93 (2009)
  [arXiv:0803.1593 [hep-th]].

\bibitem{Alles:1997qe} 
  B.~Alles, M.~D'Elia and A.~Di Giacomo,
  Phys.\ Lett.\ B {\bf 412}, 119 (1997)
  [hep-lat/9706016].
\bibitem{Gattringer:2002mr} 
  C.~Gattringer, R.~Hoffmann and S.~Schaefer,
  Phys.\ Lett.\ B {\bf 535}, 358 (2002)
  [hep-lat/0203013].
\bibitem{Lucini:2004yh} 
  B.~Lucini, M.~Teper and U.~Wenger,
  Nucl.\ Phys.\ B {\bf 715}, 461 (2005)
  [hep-lat/0401028].
\bibitem{DelDebbio:2004vxo} 
  L.~Del Debbio, H.~Panagopoulos and E.~Vicari,
  JHEP {\bf 0409}, 028 (2004)
  [hep-th/0407068].

\bibitem{Bonati:2013tt} 
  C.~Bonati, M.~D'Elia, H.~Panagopoulos and E.~Vicari,
  Phys.\ Rev.\ Lett.\  {\bf 110}, 25, 252003 (2013)
  [arXiv:1301.7640 [hep-lat]].

\bibitem{Ilgenfritz:2012wg} 
  E.~M.~Ilgenfritz and A.~Maas,
  Phys.\ Rev.\ D {\bf 86}, 114508 (2012)
  [arXiv:1210.5963 [hep-lat]].
\bibitem{Bonati:2015uga} 
  C.~Bonati,
  JHEP {\bf 1503}, 006 (2015)
  [arXiv:1501.01172 [hep-lat]].

\bibitem{Wilson:1974sk} 
  K.~G.~Wilson,
  Phys.\ Rev.\ D {\bf 10}, 2445 (1974).

\bibitem{Metropolis:1953am} 
  N.~Metropolis, A.~W.~Rosenbluth, M.~N.~Rosenbluth, A.~H.~Teller and E.~Teller,
  J.\ Chem.\ Phys.\  {\bf 21}, 1087 (1953).
\bibitem{Creutz:1980zw} 
  M.~Creutz,
  Phys.\ Rev.\ D {\bf 21}, 2308 (1980).
\bibitem{Kennedy:1985nu} 
  A.~D.~Kennedy and B.~J.~Pendleton,
  Phys.\ Lett.\  {\bf 156B}, 393 (1985).
\bibitem{Creutz:1987xi} 
  M.~Creutz,
  Phys.\ Rev.\ D {\bf 36}, 515 (1987).
\bibitem{Cabibbo:1982zn} 
  N.~Cabibbo and E.~Marinari,
  Phys.\ Lett.\  {\bf 119B}, 387 (1982).

\bibitem{Berg:1981nw} 
  B.~Berg,
  Phys.\ Lett.\ B {\bf 104}, 475 (1981).
\bibitem{Iwasaki:1983bv} 
  Y.~Iwasaki and T.~Yoshie,
  Phys.\ Lett.\ B {\bf 131}, 159 (1983).
\bibitem{Itoh:1984pr} 
  S.~Itoh, Y.~Iwasaki and T.~Yoshie,
  Phys.\ Lett.\ B {\bf 147}, 141 (1984).
\bibitem{Teper:1985rb} 
  M.~Teper,
  Phys.\ Lett.\ B {\bf 162}, 357 (1985).
\bibitem{Ilgenfritz:1985dz} 
  E.~M.~Ilgenfritz, M.~L.~Laursen, G.~Schierholz, M.~Muller-Preussker and H.~Schiller,
  Nucl.\ Phys.\ B {\bf 268}, 693 (1986).

\bibitem{Bonati:2014tqa} 
  C.~Bonati and M.~D'Elia,
  Phys.\ Rev.\ D {\bf 89}, 105005 (2014)
  [arXiv:1401.2441 [hep-lat]].
\bibitem{Cichy:2014qta} 
  K.~Cichy, A.~Dromard, E.~Garcia-Ramos, K.~Ottnad, C.~Urbach, M.~Wagner, U.~Wenger and F.~Zimmermann,
  PoS LATTICE {\bf 2014}, 075 (2014)
  [arXiv:1411.1205 [hep-lat]].
\bibitem{Namekawa:2015wua} 
  Y.~Namekawa,
  PoS LATTICE {\bf 2014}, 344 (2015)
  [arXiv:1501.06295 [hep-lat]].
\bibitem{Alexandrou:2015yba} 
  C.~Alexandrou, A.~Athenodorou and K.~Jansen,
  Phys.\ Rev.\ D {\bf 92}, 125014 (2015)
  [arXiv:1509.04259 [hep-lat]].
\bibitem{Alexandrou:2017hqw} 
  C.~Alexandrou, A.~Athenodorou, K.~Cichy, A.~Dromard, E.~Garcia-Ramos, K.~Jansen, U.~Wenger and F.~Zimmermann,
  arXiv:1708.00696 [hep-lat].
\bibitem{Berg:2016wfw} 
  B.~A.~Berg and D.~A.~Clarke,
  Phys.\ Rev.\ D {\bf 95}, 094508 (2017)
  [arXiv:1612.07347 [hep-lat]].

\bibitem{DiVecchia:1981qi} 
  P.~Di Vecchia, K.~Fabricius, G.~C.~Rossi and G.~Veneziano,
  Nucl.\ Phys.\ B {\bf 192}, 392 (1981).
\bibitem{DiVecchia:1981hh} 
  P.~Di Vecchia, K.~Fabricius, G.~C.~Rossi and G.~Veneziano,
  Phys.\ Lett.\ B {\bf 108}, 323 (1982).

\bibitem{Panagopoulos:2011rb} 
  H.~Panagopoulos and E.~Vicari,
  JHEP {\bf 1111}, 119 (2011)
  [arXiv:1109.6815 [hep-lat]].
\bibitem{Bonati:2015sqt} 
  C.~Bonati, M.~D'Elia and A.~Scapellato,
  Phys.\ Rev.\ D {\bf 93}, 025028 (2016)
  [arXiv:1512.01544 [hep-lat]].
\bibitem{Bonati:2016tvi} 
  C.~Bonati, M.~D'Elia, P.~Rossi and E.~Vicari,
  Phys.\ Rev.\ D {\bf 94}, 085017 (2016)
  [arXiv:1607.06360 [hep-lat]].

\bibitem{Campostrini:1988cy} 
  M.~Campostrini, A.~Di Giacomo and H.~Panagopoulos,
  Phys.\ Lett.\ B {\bf 212}, 206 (1988).

\bibitem{Sommer:1993ce} 
  R.~Sommer,
  Nucl.\ Phys.\ B {\bf 411}, 839 (1994)
  [hep-lat/9310022].
\bibitem{Guagnelli:1998ud} 
  M.~Guagnelli {\it et al.} [ALPHA Collaboration],
  Nucl.\ Phys.\ B {\bf 535}, 389 (1998)
  [hep-lat/9806005].

\bibitem{Luscher:2010iy} 
  M.~L\"uscher,
  JHEP {\bf 1008}, 071 (2010)
  Erratum: [JHEP {\bf 1403}, 092 (2014)]
  [arXiv:1006.4518 [hep-lat]].
\bibitem{Sommer:2014mea} 
  R.~Sommer,
  PoS LATTICE {\bf 2013}, 015 (2014)
  [arXiv:1401.3270 [hep-lat]].

\bibitem{DelDebbio:2002xa} 
  L.~Del Debbio, H.~Panagopoulos and E.~Vicari,
  JHEP {\bf 0208}, 044 (2002)
  [hep-th/0204125].

\bibitem{DElia:2003zne} 
  M.~D'Elia,
  Nucl.\ Phys.\ B {\bf 661}, 139 (2003)
  [hep-lat/0302007].

\bibitem{Giusti:2007tu} 
  L.~Giusti, S.~Petrarca and B.~Taglienti,
  Phys.\ Rev.\ D {\bf 76}, 094510 (2007)
  [arXiv:0705.2352 [hep-th]].

\bibitem{Ce:2015qha} 
M.~C\`e, C.~Consonni, G.~P.~Engel and L.~Giusti,
  Phys.\ Rev.\ D {\bf 92}, no. 7, 074502 (2015)
  [arXiv:1506.06052 [hep-lat]].


\bibitem{Thomas:2011ee} 
  E.~Thomas and A.~R.~Zhitnitsky,
  Phys.\ Rev.\ D {\bf 85}, 044039 (2012)
  [arXiv:1109.2608 [hep-th]].
\bibitem{Aitken:2018mbb} 
  K.~Aitken, A.~Cherman and M.~\"Unsal,
  arXiv:1804.06848 [hep-th].

\end{thebibliography}
\end{document}